\documentclass[11pt]{article}
\usepackage[utf8]{inputenc}
\usepackage[T1]{fontenc}
\usepackage{graphicx}
\usepackage{longtable}
\usepackage{wrapfig}
\usepackage{rotating}
\usepackage[normalem]{ulem}
\usepackage{amsmath}
\usepackage{amssymb}
\usepackage{capt-of}
\usepackage{hyperref}
\usepackage{authblk}
\usepackage{algorithm}
\usepackage{setspace}
\usepackage{upgreek}
\usepackage{cite}
\usepackage{algpseudocode}
\usepackage{tikz}
\usepackage{fancyhdr}
\usepackage{float}
\newcommand{\version}{\small Version 0.0.7}

\begin{document}

\title{Monadring: A lightweight consensus protocol to offer Validation-as-a-Service to AVS nodes}
\author{Yu Zhang\qquad Xiao Yan\qquad Gang Tang\qquad Helena Wang}
\affil{Verisense Network\\{\small team@verisense.network}}
\date{\small \today\\\version}
\maketitle

\begin{abstract}

Existing blockchain networks are often large-scale, requiring transactions to be synchronized across the entire network to reach consensus. On-chain computations can be prohibitively expensive, making many CPU-intensive computations infeasible. Inspired by the structure of IBM’s token ring networks, we propose a lightweight consensus protocol called Monadring to address these issues. Monadring allows nodes within a large blockchain network to form smaller subnetworks, enabling faster and more cost-effective computations while maintaining the security guarantees of	the	main	blockchain	network.

To further enhance Monadring's security, we introduce a node rotation mechanism based on Verifiable Random Function (VRF) and blind voting using Fully Homomorphic Encryption (FHE) within the smaller subnetwork. Unlike the common voting-based election of validator nodes, Monadring leverages FHE to conceal voting information, eliminating the advantage of the  last  mover  in  the  voting  process.

This paper details the design and implementation of the Monadring protocol and evaluates its performance and feasibility through simulation experiments. Our research contributes to enhancing the practical utility of blockchain technology in large-scale application scenarios.

\end{abstract}

\section{Introduction}
Recent blockchain systems have adopted faster and more energy-efficient consensus protocols like Ouroboros\cite{cryptoeprint:2016/889}, BABE\cite{burdges2020overview}, and Tendermint\cite{buchman2016tendermint}. These networks are typically open and permissionless, with their degree of decentralization defined by the number of participating nodes. Leveraging the aforementioned consensus protocols, new blockchain networks can be rapidly constructed. However, decentralized applications often cannot flexibly customize their degree of decentralization based on the importance of their underlying data.

For instance, decentralized social media applications may require faster response times and lower storage costs compared to decentralized finance applications, necessitating a lower degree of decentralization. This flexibility to adjust the level of decentralization based on the needs of the application is a critical consideration that existing blockchain architectures have not fully addressed.

Building a subnetwork over an existing blockchain system is a potential solution to this issue. Inspired by the token ring architecture, we propose a novel consensus protocol called Monadring. Monadring enables nodes within an existing blockchain network to form smaller, lightweight subnetworks that can perform computations more efficiently and cost-effectively while maintaining the same security guarantees as the main blockchain.

The token ring network operates at the data link layer of the OSI/RM model and was designed to solve the problem of physical link contention, similar to Ethernet. In a token ring network, a token is passed sequentially from one node to the next, granting the holder the right to transmit data. This token-based system for managing access to the shared medium bears some similarity to the consensus mechanisms used in blockchain networks for selecting block producers.

Despite token ring networks largely falling out of favor in modern networking due to their limited scalability and other drawbacks, the underlying principles of their decentralized, token-based structure offer valuable insights for designing a lightweight consensus algorithm for small blockchain systems.

Monadring also involves an FHE-based member rotation mechanism to enhance security. Homomorphic encryption (HE) \cite{acar2018survey} is a method of encryption that allows computations to be carried out on encrypted data, generating an encrypted result which, when decrypted, matches the outcome of computations performed on the plaintext. This property enables sophisticated computations on encrypted data while maintaining data security. HE schemes protect data privacy by allowing computations to be performed directly on encrypted data. For example, an HE scheme might allow a user to perform operations like addition and multiplication on encrypted numbers, with the same result as if they were performed on the original, unencrypted numbers. This technology is essential for secure cloud computing as it allows complex data manipulations on completely secure encrypted data.

Fully Homomorphic Encryption (FHE)\cite{chillotti2020tfhe} is a more advanced form of Homomorphic Encryption. FHE allows arbitrary computations to be carried out on encrypted data, unlike normal HE, which may be limited in the types of computations it supports. FHE computations generate a result that, when decrypted, corresponds to the result of the same computations performed on the plaintext. This makes FHE extremely useful for scenarios where sensitive data must be processed or analyzed, but security and privacy considerations prevent the data from being decrypted. With FHE, unlimited calculations can be performed on encrypted data just as on unencrypted data. For instance, in cloud computing, FHE allows users to operate computations on encrypted data stored in the cloud, preserving data confidentiality and privacy.

In small-scale decentralized networks with randomly determined members, the application of FHE for voting can be mathematically equivalent to the Prisoner’s Dilemma. By incorporating appropriate incentive structures, it is possible to achieve both efficiency and security in such decentralized networks.

\subsection{Overall Framework}
The framework is illustrated as Fig.~\ref{fig:framework}.
Section \ref{sec:math} demonstrates the mathematical model for transforming perfect information games in voting into imperfect information games through FHE, thereby eliminating the advantage of the last voter. Section \ref{sec:protocol} outlines the Monadring protocol. The topological structure of Monadring and Token Circulation are detailed in Section \ref{sec:subnet_topology}, while initialization and Node Rotation are explained in Section \ref{sec:launch_subnet_and_rotate}. Fault Tolerance mechanisms are discussed in Section \ref{sec:fault_tolerance}.  Section \ref{sec:key_sharing} covers the processes of Key Sharing and Key Resharing.

\tikzset{every picture/.style={line width=0.75pt}} 
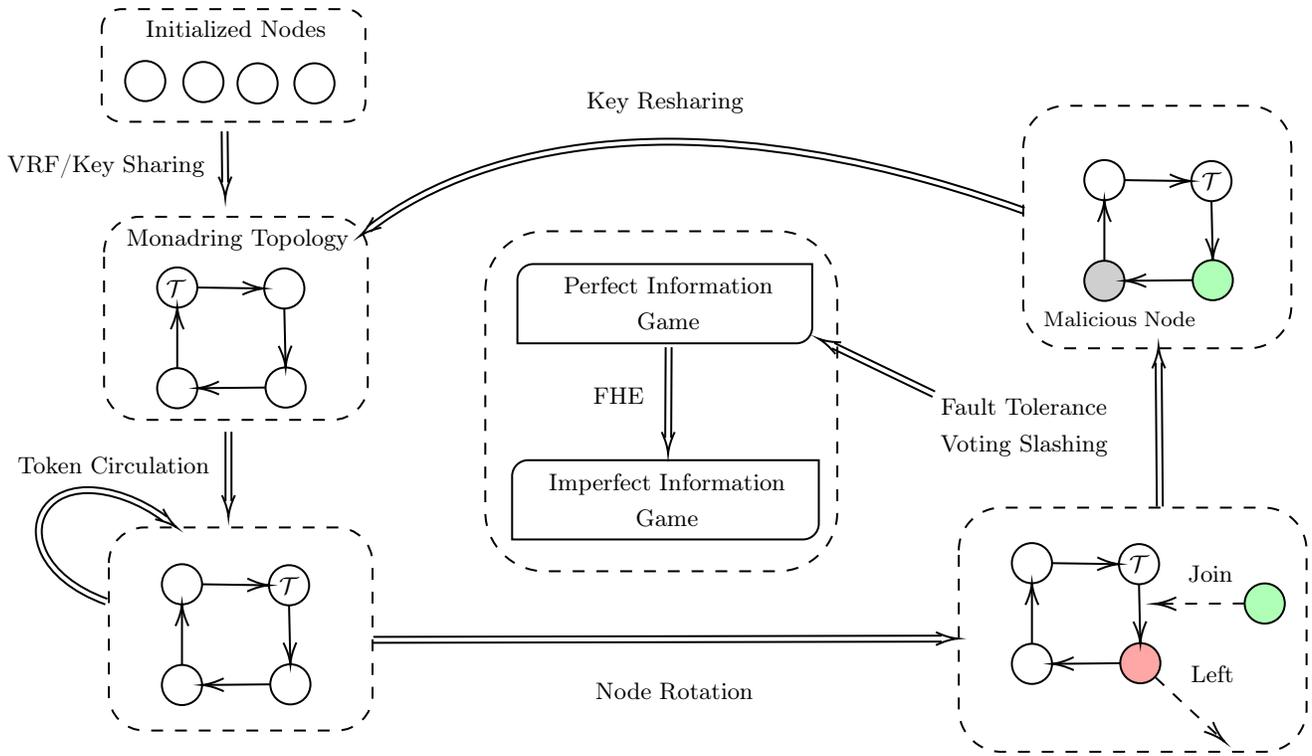
\begin{figure}[H]
  \centering
  \hspace*{-80pt}
\begin{tikzpicture}[x=0.75pt,y=0.75pt,yscale=-1,xscale=1]
  
  \draw   (68.17,52.96) .. controls (68.17,47.37) and (72.7,42.83) .. (78.29,42.83) .. controls (83.88,42.83) and (88.42,47.37) .. (88.42,52.96) .. controls (88.42,58.55) and (83.88,63.08) .. (78.29,63.08) .. controls (72.7,63.08) and (68.17,58.55) .. (68.17,52.96) -- cycle ;
  \draw   (97.5,53.46) .. controls (97.5,47.87) and (102.03,43.33) .. (107.63,43.33) .. controls (113.22,43.33) and (117.75,47.87) .. (117.75,53.46) .. controls (117.75,59.05) and (113.22,63.58) .. (107.63,63.58) .. controls (102.03,63.58) and (97.5,59.05) .. (97.5,53.46) -- cycle ;
  \draw   (124.83,54.13) .. controls (124.83,48.53) and (129.37,44) .. (134.96,44) .. controls (140.55,44) and (145.08,48.53) .. (145.08,54.13) .. controls (145.08,59.72) and (140.55,64.25) .. (134.96,64.25) .. controls (129.37,64.25) and (124.83,59.72) .. (124.83,54.13) -- cycle ;
  \draw   (153.5,54.12) .. controls (153.5,48.53) and (158.03,44) .. (163.63,44) .. controls (169.22,44) and (173.75,48.53) .. (173.75,54.12) .. controls (173.75,59.72) and (169.22,64.25) .. (163.63,64.25) .. controls (158.03,64.25) and (153.5,59.72) .. (153.5,54.12) -- cycle ;
  \draw    (119.83,78.32) -- (120.09,104.16)(116.83,78.35) -- (117.09,104.19) ;
  \draw [shift={(118.67,112.17)}, rotate = 269.44] [color={rgb, 255:red, 0; green, 0; blue, 0 }  ][line width=0.75]    (10.93,-3.29) .. controls (6.95,-1.4) and (3.31,-0.3) .. (0,0) .. controls (3.31,0.3) and (6.95,1.4) .. (10.93,3.29)   ;
  \draw  [color={rgb, 255:red, 0; green, 0; blue, 0 }  ,draw opacity=1 ] (84.3,157.16) .. controls (84.3,151.57) and (88.83,147.03) .. (94.43,147.03) .. controls (100.02,147.03) and (104.55,151.57) .. (104.55,157.16) .. controls (104.55,162.75) and (100.02,167.28) .. (94.43,167.28) .. controls (88.83,167.28) and (84.3,162.75) .. (84.3,157.16) -- cycle ;
  \draw   (84.3,207.99) .. controls (84.3,202.4) and (88.83,197.87) .. (94.43,197.87) .. controls (100.02,197.87) and (104.55,202.4) .. (104.55,207.99) .. controls (104.55,213.58) and (100.02,218.12) .. (94.43,218.12) .. controls (88.83,218.12) and (84.3,213.58) .. (84.3,207.99) -- cycle ;
  \draw   (138.97,207.66) .. controls (138.97,202.07) and (143.5,197.53) .. (149.09,197.53) .. controls (154.68,197.53) and (159.22,202.07) .. (159.22,207.66) .. controls (159.22,213.25) and (154.68,217.78) .. (149.09,217.78) .. controls (143.5,217.78) and (138.97,213.25) .. (138.97,207.66) -- cycle ;
  \draw   (138.3,157.66) .. controls (138.3,152.07) and (142.83,147.53) .. (148.43,147.53) .. controls (154.02,147.53) and (158.55,152.07) .. (158.55,157.66) .. controls (158.55,163.25) and (154.02,167.78) .. (148.43,167.78) .. controls (142.83,167.78) and (138.3,163.25) .. (138.3,157.66) -- cycle ;
  \draw  [dash pattern={on 4.5pt off 4.5pt}] (56.33,26.61) .. controls (56.33,21.12) and (60.79,16.67) .. (66.28,16.67) -- (179.39,16.67) .. controls (184.88,16.67) and (189.33,21.12) .. (189.33,26.61) -- (189.33,63.56) .. controls (189.33,69.05) and (184.88,73.51) .. (179.39,73.51) -- (66.28,73.51) .. controls (60.79,73.51) and (56.33,69.05) .. (56.33,63.56) -- cycle ;
  \draw    (104.55,157.16) -- (136.3,157.63) ;
  \draw [shift={(138.3,157.66)}, rotate = 180.85] [color={rgb, 255:red, 0; green, 0; blue, 0 }  ][line width=0.75]    (10.93,-3.29) .. controls (6.95,-1.4) and (3.31,-0.3) .. (0,0) .. controls (3.31,0.3) and (6.95,1.4) .. (10.93,3.29)   ;
  \draw    (148.43,167.78) -- (149.05,195.53) ;
  \draw [shift={(149.09,197.53)}, rotate = 268.72] [color={rgb, 255:red, 0; green, 0; blue, 0 }  ][line width=0.75]    (10.93,-3.29) .. controls (6.95,-1.4) and (3.31,-0.3) .. (0,0) .. controls (3.31,0.3) and (6.95,1.4) .. (10.93,3.29)   ;
  \draw    (138.97,207.66) -- (106.55,207.97) ;
  \draw [shift={(104.55,207.99)}, rotate = 359.45] [color={rgb, 255:red, 0; green, 0; blue, 0 }  ][line width=0.75]    (10.93,-3.29) .. controls (6.95,-1.4) and (3.31,-0.3) .. (0,0) .. controls (3.31,0.3) and (6.95,1.4) .. (10.93,3.29)   ;
  \draw    (94.43,197.87) -- (94.43,169.28) ;
  \draw [shift={(94.43,167.28)}, rotate = 90] [color={rgb, 255:red, 0; green, 0; blue, 0 }  ][line width=0.75]    (10.93,-3.29) .. controls (6.95,-1.4) and (3.31,-0.3) .. (0,0) .. controls (3.31,0.3) and (6.95,1.4) .. (10.93,3.29)   ;
  \draw  [dash pattern={on 4.5pt off 4.5pt}] (57.53,139.41) .. controls (57.53,129.5) and (65.56,121.47) .. (75.47,121.47) -- (172.59,121.47) .. controls (182.5,121.47) and (190.53,129.5) .. (190.53,139.41) -- (190.53,206.03) .. controls (190.53,215.94) and (182.5,223.97) .. (172.59,223.97) -- (75.47,223.97) .. controls (65.56,223.97) and (57.53,215.94) .. (57.53,206.03) -- cycle ;
  \draw    (121.77,229.97) -- (121.77,264.97)(118.77,229.97) -- (118.77,264.97) ;
  \draw [shift={(120.27,272.97)}, rotate = 270] [color={rgb, 255:red, 0; green, 0; blue, 0 }  ][line width=0.75]    (10.93,-3.29) .. controls (6.95,-1.4) and (3.31,-0.3) .. (0,0) .. controls (3.31,0.3) and (6.95,1.4) .. (10.93,3.29)   ;
  \draw  [color={rgb, 255:red, 0; green, 0; blue, 0 }  ,draw opacity=1 ] (86.7,306.96) .. controls (86.7,301.37) and (91.23,296.83) .. (96.82,296.83) .. controls (102.42,296.83) and (106.95,301.37) .. (106.95,306.96) .. controls (106.95,312.55) and (102.42,317.08) .. (96.82,317.08) .. controls (91.23,317.08) and (86.7,312.55) .. (86.7,306.96) -- cycle ;
  \draw   (86.7,357.79) .. controls (86.7,352.2) and (91.23,347.67) .. (96.82,347.67) .. controls (102.42,347.67) and (106.95,352.2) .. (106.95,357.79) .. controls (106.95,363.38) and (102.42,367.92) .. (96.82,367.92) .. controls (91.23,367.92) and (86.7,363.38) .. (86.7,357.79) -- cycle ;
  \draw   (141.37,357.46) .. controls (141.37,351.87) and (145.9,347.33) .. (151.49,347.33) .. controls (157.08,347.33) and (161.62,351.87) .. (161.62,357.46) .. controls (161.62,363.05) and (157.08,367.58) .. (151.49,367.58) .. controls (145.9,367.58) and (141.37,363.05) .. (141.37,357.46) -- cycle ;
  \draw   (140.7,307.46) .. controls (140.7,301.87) and (145.23,297.33) .. (150.82,297.33) .. controls (156.42,297.33) and (160.95,301.87) .. (160.95,307.46) .. controls (160.95,313.05) and (156.42,317.58) .. (150.82,317.58) .. controls (145.23,317.58) and (140.7,313.05) .. (140.7,307.46) -- cycle ;
  \draw    (106.95,306.96) -- (138.7,307.43) ;
  \draw [shift={(140.7,307.46)}, rotate = 180.85] [color={rgb, 255:red, 0; green, 0; blue, 0 }  ][line width=0.75]    (10.93,-3.29) .. controls (6.95,-1.4) and (3.31,-0.3) .. (0,0) .. controls (3.31,0.3) and (6.95,1.4) .. (10.93,3.29)   ;
  \draw    (150.82,317.58) -- (151.45,345.33) ;
  \draw [shift={(151.49,347.33)}, rotate = 268.72] [color={rgb, 255:red, 0; green, 0; blue, 0 }  ][line width=0.75]    (10.93,-3.29) .. controls (6.95,-1.4) and (3.31,-0.3) .. (0,0) .. controls (3.31,0.3) and (6.95,1.4) .. (10.93,3.29)   ;
  \draw    (141.37,357.46) -- (108.95,357.77) ;
  \draw [shift={(106.95,357.79)}, rotate = 359.45] [color={rgb, 255:red, 0; green, 0; blue, 0 }  ][line width=0.75]    (10.93,-3.29) .. controls (6.95,-1.4) and (3.31,-0.3) .. (0,0) .. controls (3.31,0.3) and (6.95,1.4) .. (10.93,3.29)   ;
  \draw    (96.82,347.67) -- (96.82,319.08) ;
  \draw [shift={(96.82,317.08)}, rotate = 90] [color={rgb, 255:red, 0; green, 0; blue, 0 }  ][line width=0.75]    (10.93,-3.29) .. controls (6.95,-1.4) and (3.31,-0.3) .. (0,0) .. controls (3.31,0.3) and (6.95,1.4) .. (10.93,3.29)   ;
  \draw  [dash pattern={on 4.5pt off 4.5pt}] (59.93,296.21) .. controls (59.93,286.3) and (67.96,278.27) .. (77.87,278.27) -- (174.99,278.27) .. controls (184.9,278.27) and (192.93,286.3) .. (192.93,296.21) -- (192.93,362.83) .. controls (192.93,372.74) and (184.9,380.77) .. (174.99,380.77) -- (77.87,380.77) .. controls (67.96,380.77) and (59.93,372.74) .. (59.93,362.83) -- cycle ;
  \draw    (58.15,317.12) .. controls (40.83,310.79) and (30.56,301.38) .. (25.99,292.03) .. controls (24.04,288.04) and (23.11,284.04) .. (23.11,280.26) .. controls (23.11,272.59) and (26.91,265.81) .. (33.79,261.78) .. controls (37.92,259.36) and (43.17,257.93) .. (49.39,257.93) .. controls (61.08,257.93) and (76.32,263.01) .. (89.22,272.85)(59.18,314.3) .. controls (42.84,308.33) and (33.01,299.56) .. (28.69,290.72) .. controls (26.95,287.17) and (26.11,283.62) .. (26.11,280.26) .. controls (26.11,273.66) and (29.39,267.84) .. (35.31,264.37) .. controls (39.03,262.18) and (43.78,260.93) .. (49.39,260.93) .. controls (60.63,260.93) and (75.23,265.95) .. (87.36,275.21) ;
  \draw [shift={(94,278.17)}, rotate = 217.4] [color={rgb, 255:red, 0; green, 0; blue, 0 }  ][line width=0.75]    (10.93,-3.29) .. controls (6.95,-1.4) and (3.31,-0.3) .. (0,0) .. controls (3.31,0.3) and (6.95,1.4) .. (10.93,3.29)   ;
  \draw    (193.33,333.5) -- (480,332.89)(193.34,336.5) -- (480,335.89) ;
  \draw [shift={(488,334.38)}, rotate = 179.88] [color={rgb, 255:red, 0; green, 0; blue, 0 }  ][line width=0.75]    (10.93,-3.29) .. controls (6.95,-1.4) and (3.31,-0.3) .. (0,0) .. controls (3.31,0.3) and (6.95,1.4) .. (10.93,3.29)   ;
  \draw  [color={rgb, 255:red, 0; green, 0; blue, 0 }  ,draw opacity=1 ] (515.37,296.29) .. controls (515.37,290.7) and (519.9,286.17) .. (525.49,286.17) .. controls (531.08,286.17) and (535.62,290.7) .. (535.62,296.29) .. controls (535.62,301.88) and (531.08,306.42) .. (525.49,306.42) .. controls (519.9,306.42) and (515.37,301.88) .. (515.37,296.29) -- cycle ;
  \draw   (515.37,347.13) .. controls (515.37,341.53) and (519.9,337) .. (525.49,337) .. controls (531.08,337) and (535.62,341.53) .. (535.62,347.13) .. controls (535.62,352.72) and (531.08,357.25) .. (525.49,357.25) .. controls (519.9,357.25) and (515.37,352.72) .. (515.37,347.13) -- cycle ;
  \draw  [fill={rgb, 255:red, 255; green, 166; blue, 166 }  ,fill opacity=1 ] (570.03,346.79) .. controls (570.03,341.2) and (574.57,336.67) .. (580.16,336.67) .. controls (585.75,336.67) and (590.28,341.2) .. (590.28,346.79) .. controls (590.28,352.38) and (585.75,356.92) .. (580.16,356.92) .. controls (574.57,356.92) and (570.03,352.38) .. (570.03,346.79) -- cycle ;
  \draw   (569.37,296.79) .. controls (569.37,291.2) and (573.9,286.67) .. (579.49,286.67) .. controls (585.08,286.67) and (589.62,291.2) .. (589.62,296.79) .. controls (589.62,302.38) and (585.08,306.92) .. (579.49,306.92) .. controls (573.9,306.92) and (569.37,302.38) .. (569.37,296.79) -- cycle ;
  \draw    (535.62,296.29) -- (567.37,296.76) ;
  \draw [shift={(569.37,296.79)}, rotate = 180.85] [color={rgb, 255:red, 0; green, 0; blue, 0 }  ][line width=0.75]    (10.93,-3.29) .. controls (6.95,-1.4) and (3.31,-0.3) .. (0,0) .. controls (3.31,0.3) and (6.95,1.4) .. (10.93,3.29)   ;
  \draw    (579.49,306.92) -- (580.11,334.67) ;
  \draw [shift={(580.16,336.67)}, rotate = 268.72] [color={rgb, 255:red, 0; green, 0; blue, 0 }  ][line width=0.75]    (10.93,-3.29) .. controls (6.95,-1.4) and (3.31,-0.3) .. (0,0) .. controls (3.31,0.3) and (6.95,1.4) .. (10.93,3.29)   ;
  \draw    (570.03,346.79) -- (537.62,347.11) ;
  \draw [shift={(535.62,347.13)}, rotate = 359.45] [color={rgb, 255:red, 0; green, 0; blue, 0 }  ][line width=0.75]    (10.93,-3.29) .. controls (6.95,-1.4) and (3.31,-0.3) .. (0,0) .. controls (3.31,0.3) and (6.95,1.4) .. (10.93,3.29)   ;
  \draw    (525.49,337) -- (525.49,308.42) ;
  \draw [shift={(525.49,306.42)}, rotate = 90] [color={rgb, 255:red, 0; green, 0; blue, 0 }  ][line width=0.75]    (10.93,-3.29) .. controls (6.95,-1.4) and (3.31,-0.3) .. (0,0) .. controls (3.31,0.3) and (6.95,1.4) .. (10.93,3.29)   ;
  \draw  [dash pattern={on 4.5pt off 4.5pt}] (488.6,289.83) .. controls (488.6,277.92) and (498.26,268.27) .. (510.17,268.27) -- (642.43,268.27) .. controls (654.34,268.27) and (664,277.92) .. (664,289.83) -- (664,369.94) .. controls (664,381.85) and (654.34,391.51) .. (642.43,391.51) -- (510.17,391.51) .. controls (498.26,391.51) and (488.6,381.85) .. (488.6,369.94) -- cycle ;
  \draw  [fill={rgb, 255:red, 175; green, 255; blue, 182 }  ,fill opacity=1 ] (632.77,316.66) .. controls (632.77,311.07) and (637.3,306.53) .. (642.89,306.53) .. controls (648.48,306.53) and (653.02,311.07) .. (653.02,316.66) .. controls (653.02,322.25) and (648.48,326.78) .. (642.89,326.78) .. controls (637.3,326.78) and (632.77,322.25) .. (632.77,316.66) -- cycle ;
  \draw  [dash pattern={on 4.5pt off 4.5pt}]  (632.77,316.66) -- (589.33,316.83) ;
  \draw [shift={(587.33,316.84)}, rotate = 359.77] [color={rgb, 255:red, 0; green, 0; blue, 0 }  ][line width=0.75]    (10.93,-3.29) .. controls (6.95,-1.4) and (3.31,-0.3) .. (0,0) .. controls (3.31,0.3) and (6.95,1.4) .. (10.93,3.29)   ;
  \draw  [dash pattern={on 4.5pt off 4.5pt}]  (587.8,353.97) -- (621.14,385.01) ;
  \draw [shift={(622.6,386.37)}, rotate = 222.95] [color={rgb, 255:red, 0; green, 0; blue, 0 }  ][line width=0.75]    (10.93,-3.29) .. controls (6.95,-1.4) and (3.31,-0.3) .. (0,0) .. controls (3.31,0.3) and (6.95,1.4) .. (10.93,3.29)   ;
  \draw  [color={rgb, 255:red, 0; green, 0; blue, 0 }  ,draw opacity=1 ] (551.9,102.96) .. controls (551.9,97.37) and (556.43,92.83) .. (562.02,92.83) .. controls (567.62,92.83) and (572.15,97.37) .. (572.15,102.96) .. controls (572.15,108.55) and (567.62,113.08) .. (562.02,113.08) .. controls (556.43,113.08) and (551.9,108.55) .. (551.9,102.96) -- cycle ;
  \draw  [fill={rgb, 255:red, 207; green, 207; blue, 207 }  ,fill opacity=1 ] (551.9,153.79) .. controls (551.9,148.2) and (556.43,143.67) .. (562.02,143.67) .. controls (567.62,143.67) and (572.15,148.2) .. (572.15,153.79) .. controls (572.15,159.38) and (567.62,163.92) .. (562.02,163.92) .. controls (556.43,163.92) and (551.9,159.38) .. (551.9,153.79) -- cycle ;
  \draw   (605.9,103.46) .. controls (605.9,97.87) and (610.43,93.33) .. (616.02,93.33) .. controls (621.62,93.33) and (626.15,97.87) .. (626.15,103.46) .. controls (626.15,109.05) and (621.62,113.58) .. (616.02,113.58) .. controls (610.43,113.58) and (605.9,109.05) .. (605.9,103.46) -- cycle ;
  \draw    (572.15,102.96) -- (603.9,103.43) ;
  \draw [shift={(605.9,103.46)}, rotate = 180.85] [color={rgb, 255:red, 0; green, 0; blue, 0 }  ][line width=0.75]    (10.93,-3.29) .. controls (6.95,-1.4) and (3.31,-0.3) .. (0,0) .. controls (3.31,0.3) and (6.95,1.4) .. (10.93,3.29)   ;
  \draw    (616.02,113.58) -- (616.65,141.33) ;
  \draw [shift={(616.69,143.33)}, rotate = 268.72] [color={rgb, 255:red, 0; green, 0; blue, 0 }  ][line width=0.75]    (10.93,-3.29) .. controls (6.95,-1.4) and (3.31,-0.3) .. (0,0) .. controls (3.31,0.3) and (6.95,1.4) .. (10.93,3.29)   ;
  \draw    (606.57,153.46) -- (574.15,153.77) ;
  \draw [shift={(572.15,153.79)}, rotate = 359.45] [color={rgb, 255:red, 0; green, 0; blue, 0 }  ][line width=0.75]    (10.93,-3.29) .. controls (6.95,-1.4) and (3.31,-0.3) .. (0,0) .. controls (3.31,0.3) and (6.95,1.4) .. (10.93,3.29)   ;
  \draw    (562.02,143.67) -- (562.02,115.08) ;
  \draw [shift={(562.02,113.08)}, rotate = 90] [color={rgb, 255:red, 0; green, 0; blue, 0 }  ][line width=0.75]    (10.93,-3.29) .. controls (6.95,-1.4) and (3.31,-0.3) .. (0,0) .. controls (3.31,0.3) and (6.95,1.4) .. (10.93,3.29)   ;
  \draw  [dash pattern={on 4.5pt off 4.5pt}] (521.13,86.86) .. controls (521.13,75.03) and (530.72,65.44) .. (542.55,65.44) -- (634.91,65.44) .. controls (646.74,65.44) and (656.33,75.03) .. (656.33,86.86) -- (656.33,166.42) .. controls (656.33,178.25) and (646.74,187.84) .. (634.91,187.84) -- (542.55,187.84) .. controls (530.72,187.84) and (521.13,178.25) .. (521.13,166.42) -- cycle ;
  \draw  [fill={rgb, 255:red, 175; green, 255; blue, 182 }  ,fill opacity=1 ] (606.57,153.46) .. controls (606.57,147.87) and (611.1,143.33) .. (616.69,143.33) .. controls (622.28,143.33) and (626.82,147.87) .. (626.82,153.46) .. controls (626.82,159.05) and (622.28,163.58) .. (616.69,163.58) .. controls (611.1,163.58) and (606.57,159.05) .. (606.57,153.46) -- cycle ;
  \draw    (588.5,267.72) -- (587.9,195.05)(591.5,267.7) -- (590.9,195.03) ;
  \draw [shift={(589.33,187.04)}, rotate = 89.53] [color={rgb, 255:red, 0; green, 0; blue, 0 }  ][line width=0.75]    (10.93,-3.29) .. controls (6.95,-1.4) and (3.31,-0.3) .. (0,0) .. controls (3.31,0.3) and (6.95,1.4) .. (10.93,3.29)   ;
  \draw    (520.83,119.79) .. controls (510.72,116.21) and (493.36,110.12) .. (471.23,104.03) .. controls (436.8,94.55) and (390.8,85.07) .. (342.55,85.07) .. controls (330.9,85.07) and (319.12,85.63) .. (307.34,86.87) .. controls (265.45,91.28) and (223.55,104.37) .. (191.93,128.96)(521.83,116.96) .. controls (511.68,113.37) and (494.25,107.25) .. (472.02,101.14) .. controls (437.38,91.6) and (391.1,82.07) .. (342.55,82.07) .. controls (330.8,82.07) and (318.91,82.63) .. (307.03,83.88) .. controls (264.59,88.35) and (222.17,101.65) .. (190.15,126.54) ;
  \draw [shift={(186.67,131.04)}, rotate = 322.24] [color={rgb, 255:red, 0; green, 0; blue, 0 }  ][line width=0.75]    (10.93,-3.29) .. controls (6.95,-1.4) and (3.31,-0.3) .. (0,0) .. controls (3.31,0.3) and (6.95,1.4) .. (10.93,3.29)   ;
  \draw   (266,153.33) .. controls (266,148.92) and (269.58,145.33) .. (274,145.33) -- (414.67,145.33) .. controls (414.67,145.33) and (414.67,145.33) .. (414.67,145.33) -- (414.67,177.33) .. controls (414.67,181.75) and (411.08,185.33) .. (406.67,185.33) -- (266,185.33) .. controls (266,185.33) and (266,185.33) .. (266,185.33) -- cycle ;
  \draw   (263.33,252.33) .. controls (263.33,247.92) and (266.92,244.33) .. (271.33,244.33) -- (418,244.33) .. controls (418,244.33) and (418,244.33) .. (418,244.33) -- (418,276.33) .. controls (418,280.75) and (414.42,284.33) .. (410,284.33) -- (263.33,284.33) .. controls (263.33,284.33) and (263.33,284.33) .. (263.33,284.33) -- cycle ;
  \draw  [color={rgb, 255:red, 0; green, 0; blue, 0 }  ,draw opacity=1 ][dash pattern={on 4.5pt off 4.5pt}] (249.8,158.8) .. controls (249.8,142.22) and (263.25,128.77) .. (279.83,128.77) -- (397.3,128.77) .. controls (413.89,128.77) and (427.33,142.22) .. (427.33,158.8) -- (427.33,270.34) .. controls (427.33,286.93) and (413.89,300.38) .. (397.3,300.38) -- (279.83,300.38) .. controls (263.25,300.38) and (249.8,286.93) .. (249.8,270.34) -- cycle ;
  \draw    (343.77,187.31) -- (343.54,233.05)(340.77,187.3) -- (340.54,233.03) ;
  \draw [shift={(342,241.04)}, rotate = 270.28] [color={rgb, 255:red, 0; green, 0; blue, 0 }  ][line width=0.75]    (10.93,-3.29) .. controls (6.95,-1.4) and (3.31,-0.3) .. (0,0) .. controls (3.31,0.3) and (6.95,1.4) .. (10.93,3.29)   ;
  \draw    (475.35,212.35) -- (424.56,187.87)(476.65,209.65) -- (425.86,185.16) ;
  \draw [shift={(418,183.04)}, rotate = 25.74] [color={rgb, 255:red, 0; green, 0; blue, 0 }  ][line width=0.75]    (10.93,-3.29) .. controls (6.95,-1.4) and (3.31,-0.3) .. (0,0) .. controls (3.31,0.3) and (6.95,1.4) .. (10.93,3.29)   ;
  
  \draw (124.03,26.5) node  [font=\normalsize] [align=left] {{\footnotesize Initialized Nodes}};
  \draw (58.49,95.87) node  [font=\footnotesize] [align=left] {VRF/Key Sharing};
  \draw (125.13,133.1) node   [align=left] {{\footnotesize Monadring Topology}};
  \draw (93.63,157.3) node  [font=\footnotesize] [align=left] {$\displaystyle \mathcal{T}$};
  \draw (62.49,246.61) node   [align=left] {{\footnotesize Token Circulation}};
  \draw (150.82,307.46) node  [font=\footnotesize] [align=left] {$\displaystyle \mathcal{T}$};
  \draw (304,355.33) node [anchor=north west][inner sep=0.75pt]   [align=left] {{\footnotesize Node Rotation}};
  \draw (579.49,296.79) node  [font=\footnotesize] [align=left] {$\displaystyle \mathcal{T}$};
  \draw (602.73,295.87) node [anchor=north west][inner sep=0.75pt]   [align=left] {{\footnotesize Join}};
  \draw (604.2,346.57) node [anchor=north west][inner sep=0.75pt]   [align=left] {{\footnotesize Left}};
  \draw (616.02,103.46) node  [font=\footnotesize] [align=left] {$\displaystyle \mathcal{T}$};
  \draw (302.67,205.85) node [anchor=north west][inner sep=0.75pt]   [align=left] {{\footnotesize FHE}};
  \draw (299.33,57) node [anchor=north west][inner sep=0.75pt]   [align=left] {{\footnotesize Key Resharing}};
  \draw (342.27,165.02) node   [align=left] {\begin{minipage}[lt]{81.96pt}\setlength\topsep{0pt}
  \begin{center}
  {\footnotesize Perfect Information Game}
  \end{center}
  
  \end{minipage}};
  \draw (341.67,264.35) node   [align=left] {\begin{minipage}[lt]{98.37pt}\setlength\topsep{0pt}
  \begin{center}
  {\footnotesize Imperfect Information Game}
  \end{center}
  
  \end{minipage}};
  \draw (521.6,227.69) node   [align=left] {\begin{minipage}[lt]{81.96pt}\setlength\topsep{0pt}
  \begin{center}
  {\footnotesize Fault Tolerance}\\{\footnotesize Voting Slashing}
  \end{center}
  
  \end{minipage}};
  \draw (530,167.67) node [anchor=north west][inner sep=0.75pt]   [align=left] {{\scriptsize Malicious Node}};

  \end{tikzpicture}

  \caption{Illustration of Monadring Framework}
  \label{fig:framework}
\end{figure}

\noindent

\section{Essential Mathematics}
\label{sec:math}
\subsection{Fully Homomorphic Encryption (FHE)}
\label{sec:BFV}

We define the fundamental number sets: $\mathbb{Z}$ (integers), $\mathbb{Q}$ (rational numbers), $\mathbb{R}$ (real numbers), and $\mathbb{C}$ (complex numbers). The polynomial cyclotomic ring $\mathcal{R}$ is defined as:

$$\mathcal{R}:=\mathbb{Z}[X]/(X^N+1)$$

Correspondingly, $\mathcal{R}_q$ is defined as $(\mathbb{Z}/q\mathbb{Z})[X]/(X^N+1)$, where typically $N=2^n$ for some integer $n$.

We denote distributions as $\chi_{\sigma^2,\mu}$, where $\sigma^2$ is the variance and $\mu$ is the mean. The uniform ternary distribution and discrete Gaussian distribution are represented by $\chi_\mathcal{T}$ and $\chi_\mathcal{N}$ respectively. In the standard BFV encryption formula:

$$\mathbf{a}\cdot \mathbf{s} +\Delta m +\mathbf{e}$$
$$\mathbf{a}\cdot \mathbf{s} + m +t\mathbf{e}$$

We have $\mathbf{a}\sim\chi_\mathcal{T}$ and $\mathbf{e}\sim\chi_{\mathcal{N}}$. The probability density function of the discrete Gaussian distribution is given by $\rho_{\mu,\sigma^2}=e^{-\lVert\mathbf{x}-\mathbf{\mu}\rVert^2/2\sigma^2}$.

The expansion factor $\delta_\mathcal{R}$ is defined as $\lVert \mathbf{a}\cdot\mathbf{b}\rVert_\infty/(\lVert\mathbf{a}\rVert_\infty\cdot\lVert\mathbf{b}\rVert_\infty)$ where $\mathbf{a},\mathbf{b}\in \mathcal{R}$.

As described in \cite{10.1007/978-3-642-32009-5_50, cryptoeprint:2012/144}, the BFV scheme places the ciphertext in the Most Significant Digit (MSD) position, changing the noise growth trend from quadratic to linear. This significantly reduces the impact of noise during computation and eliminates the need for modulus switching, thereby enhancing computational efficiency.

The BFV scheme consists of the following algorithms:

\begin{itemize}
\item \texttt{BFV.SecretKeyGen($1^\lambda$)}: Generate a secret key $\texttt{sk}\leftarrow \chi_\mathcal{T}$.
\item \texttt{BFV.PublicKeyGen(\texttt{sk})}: Generate a public key $\texttt{pk} =(\texttt{pk}_0,\texttt{pk}_1)= \left([-\mathbf{a}\cdot\mathbf{s}+\mathbf{e}]_{q}, \mathbf{a}\right)$ where $\mathbf{e}\leftarrow \chi_\mathcal{N}$ and $\mathbf{a}\leftarrow \chi_\mathcal{T}$.
\item \texttt{BFV.Enc($\texttt{pk},\mathbf{m}$)}: For message $\mathbf{m}\in \mathcal{R}_p$, where $p<q$, $\mathbf{u}\leftarrow \chi_\mathcal{T}$ and $\mathbf{e}_0, \mathbf{e}_1\leftarrow \chi_\mathcal{N}$. The ciphertext is:
    $$\texttt{ct} = (\texttt{ct}_0,\texttt{ct}_1)=\left([\texttt{pk}_0\cdot\mathbf{u}+\Delta\cdot\mathbf{m}+\mathbf{e}_0]_{q},[\texttt{pk}_1\cdot\mathbf{u}+\mathbf{e}_1]_{q}\right)$$
\item \texttt{BFV.Dec($\texttt{sk},\texttt{ct}$)}:
    $$\mathbf{m}=\left[\left\lfloor p\cdot[\texttt{ct}_0+\texttt{ct}_1\cdot\mathbf{s}]_q/q\right\rceil\right]_t$$
\end{itemize}

It's noteworthy that the BFV scheme differs from the BGV scheme\cite{brakerski2014leveled} in ciphertext placement. In BGV, the ciphertext is placed in the Least Significant Digit (LSD):

\begin{itemize}
\item \texttt{BGV.PublicKeyGen(\texttt{sk})}: $$\texttt{pk}=([-\mathbf{a}\cdot{s}+p\mathbf{e}]_q, \mathbf{a})$$
\item \texttt{BGV.Enc($\texttt{pk},\mathbf{m}$)}: $$\texttt{ct}=([ \mathbf{m}+\mathbf{u}\cdot\texttt{pk}_0+p\mathbf{e}_0]_q,[\mathbf{u}\cdot\texttt{pk}_1+p\mathbf{e}_1]_q)$$
\end{itemize}

To evaluate the effect of encryption and decryption on noise, we consider the equation:

\begin{equation}
\texttt{ct}_0+\texttt{ct}_1\cdot{s}=\Delta\mathbf{m}+p\mathbf{v}
\end{equation}

The noise $\mathbf{v}$ can be bounded by $\lVert\mathbf{v}\rVert_\infty < (q-p r_t(q))/(2p)$ where $r_p(q)=q-p\Delta$, a value determined by the modulus $q$, the plaintext scaling factor $\Delta$, and the noise ratio $p$.

This condition is crucial because the decryption process involves a $p/q$ scaling operation, which may amplify the noise. Therefore, we must ensure that the noise magnitude remains within an acceptable range before decryption.

\subsection{Threshold Key Sharing}
\label{sec:key_sharing}

\subsubsection{Shamir's Secret Sharing Scheme}
\label{sec:shamir}

Shamir's Secret Sharing\cite{pang2005new}, devised by Adi Shamir, is a cryptographic algorithm that enables secure distribution of a secret among multiple participants. Its key feature is the minimal number of shares required to reconstruct the secret. The process for threshold private key sharing using Shamir's scheme is as follows:

\begin{enumerate}
    \item \textbf{Threshold Selection:} Define a threshold $t$, where knowledge of fewer than $t$ points reveals no information about the secret, but $t$ or more points allow secret reconstruction.
    
    \item \textbf{Polynomial Generation:} Create a random polynomial of degree $t-1$, with the constant term being the secret (private key) to be shared:
    \begin{equation}
        \mathcal{P}(\mathbf{x})=a_0+a_1\mathbf{x}+a_2\mathbf{x}^2+\ldots+a_{t-1}\mathbf{x}^{t-1}
    \end{equation}
    
    \item \textbf{Share Creation:} Evaluate the polynomial at $n$ different points to create $n$ shares, where $n$ is the total number of participants. Each participant receives one share, a point on the polynomial:
    \begin{equation}
        s_i=\mathcal{P}(\mathbf{x}_i)
    \end{equation}
    
    \item \textbf{Share Distribution:} Distribute the private key shares among the participants.
    
    \item \textbf{Secret Reconstruction:} When the private key is needed, any $t$ participants can combine their shares using polynomial interpolation (e.g., Lagrange interpolation) to reconstruct the polynomial and reveal the constant term (the secret):
    \begin{equation}
        \mathcal{P}(x)=\sum^{t-1}_{i=0}s_i\prod_{j\neq i}\frac{x-s_j}{s_i-s_j}
    \end{equation}
\end{enumerate}

It's important to note that all polynomials are defined over the ring $(\mathbb{Z}/p\mathbb{Z})[X]/X^{t}$, and Lagrange interpolation remains valid in this context.

This approach ensures that the private key is never fully revealed to any single party, and no individual can access the secret alone. It provides a balance between accessibility and security, making it particularly useful for managing risks in cryptographic key management systems.

\subsubsection{Resharing Scheme}
\label{sec:resharing}

In practical applications, it may be necessary to redistribute keys. This process, known as resharing\cite{groth2021non}, involves transforming an original $(n,t)$ key sharing scheme into a $(n',t')$ scheme, which can be achieved using Lagrange interpolation.

Assume an initial $(n,t)$ key sharing of $s$, where $s=\mathcal{P}(0)$. Let $\mathcal{L}^\mathcal{N}_i$ be the Lagrange basis for the original $(n,t)$ key sharing user set $\mathcal{N}$, and $\mathcal{M}$ be the user set for the new $(n',t')$ sharing. The resharing process can be expressed as:

\begin{equation}
\begin{aligned}
\mathcal{P}(0)&=\sum_{i\in\mathcal{N}}\mathcal{L}^\mathcal{N}_is_i=\sum_{i\in\mathcal{N}}\mathcal{L}^\mathcal{N}_i\sum_{j\in\mathcal{M}}\mathcal{L}^\mathcal{M}_js'_{i,j}\\
&=\sum_{j\in\mathcal{M}}\left(\sum_{i\in\mathcal{N}}\mathcal{L}^\mathcal{N}_i\mathcal{L}^\mathcal{M}_j\right)s'_{i,j}
=\sum_{j\in\mathcal{M}}\mathcal{L}^\mathcal{M}_j\sum_{i\in\mathcal{N}}\mathcal{L}^\mathcal{N}_is'_{i,j}=\sum_{j\in\mathcal{M}}\mathcal{L}^\mathcal{M}_js'_j
\end{aligned}
\end{equation}

Mathematically, this resharing process is analogous to transforming a one-dimensional vector into a two-dimensional matrix of size $n \times n'$, where any $t \times t'$ submatrix can reconstruct the original key. This transformation effectively shifts users from holding row vectors to holding column vectors.

The resharing scheme provides a flexible method for adjusting share distribution without exposing the original secret or requiring all participants to return their shares. This adaptability is crucial for maintaining long-term security and accommodating changes in the participant group over time.

\subsection{Game Theory}
\label{sec:game_theory}
This chapter elucidates the design of a game-theory\cite{fudenberg1991game} based security framework to ensure system safety and reliability. We begin by introducing fundamental concepts of game theory, followed by a discussion on transforming perfect information games into imperfect information games using fully homomorphic encryption (FHE) techniques. Finally, we present a security framework based on Bayesian games and demonstrate how game theory principles can be applied to design a secure voting mechanism, thereby enhancing overall system security.

\subsubsection{Basic Definitions}

In game theory, we consider a set of players $\mathcal{P} = \{1,2,\ldots,n\}$. Each player $i \in \mathcal{P}$ has a strategy space $\Sigma_i$, containing all possible strategies available to that player. A specific strategy chosen by player $i$ is denoted as $\sigma_i \in \Sigma_i$.

In multiplayer games, the strategies of other players often influence a player's decision. We denote the strategy profile of all players except player $i$ as $\sigma_{-i} \in \Sigma_{-i}$, where $\Sigma_{-i}$ represents all possible strategy combinations for players other than $i$.

Games can be classified based on information availability:

\begin{itemize}
  \item \textbf{Perfect Information Games:} All players have complete knowledge of all past actions and decisions within the game. The entire sequence of play is fully observable to every player.
  
  \item \textbf{Imperfect Information Games:} The complete sequence of previous actions and decisions is not fully observable for all players. Players may possess private information and typically have different information about past events. Decisions are made based on private and available public information.
\end{itemize}

\subsubsection{Nash Equilibrium}

The Nash equilibrium is a fundamental concept in game theory, representing a state where no player can gain an advantage by unilaterally changing their strategy, given that all other players' strategies remain fixed.\cite{daskalakis2009complexity}

The best response function of player $i$, denoted by $\texttt{BR}_i(\sigma_{-i})$, is defined as:

\begin{equation}
\texttt{BR}_i(\sigma_{-i}) = \arg\max_{\sigma_i \in \Sigma_i} \pi_i(\sigma_i,\sigma_{-i})
\end{equation}

where $\pi_i$ is player $i$'s utility or payoff function.

A strategy profile $\sigma^*$ is a Nash equilibrium if for all players $i$, $\sigma_i^* = \texttt{BR}_i(\sigma_{-i}^*)$.

In imperfect information games, achieving Nash equilibrium can be challenging. To address this, we consider the concept of $\epsilon$-equilibrium. A strategy profile $\sigma^\epsilon$ is an $\epsilon$-equilibrium if for all players $i$:

\begin{equation}
\pi_i(\sigma^\epsilon_i,\sigma^\epsilon_{-i}) \geq \pi_i(\sigma_i, \sigma^\epsilon_{-i}) - \epsilon
\end{equation}

This means that no player can gain more than $\epsilon$ by unilaterally deviating from their strategy.

\subsubsection{Bayesian Game}

A Bayesian game incorporates players' private information that affects their payoffs. Each player has a type $\theta_i \in \Theta_i$, drawn from a known probability distribution $p_i$, which determines their payoff function.

A Bayesian game is defined by the quintuple:

\begin{equation}
\Gamma = \langle \mathcal{P}, \{\Theta_i\}_{i\in\mathcal{P}}, \{\Sigma_i\}_{i\in\mathcal{P}}, \{\pi_i\}_{i\in\mathcal{P}}, \{p_i\}_{i\in\mathcal{P}} \rangle
\end{equation}

The Bayesian Nash Equilibrium (BNE) extends the Nash equilibrium concept to Bayesian games. A strategy profile $\sigma^*$ is a BNE if for all players $i$ and all types $\theta_i$, $\sigma_i^*(\theta_i)$ is a best response to the strategies of all other players, given $\theta_i$.

The best response function in a Bayesian game is defined as:

\begin{equation}
\sigma^\texttt{BNE}_i = \texttt{BR}^\texttt{BNE}_i(\theta_{-i},\sigma_{-i}) = \arg\max_{\substack{\sigma_i \in \Sigma_i \\ \theta_i\in \Theta_i}} \int_{\Theta_i} \pi_i(\sigma_i,\sigma^\texttt{BNE}_{-i},\theta_i,\theta_{-i})dP_i(\theta_{-i}|\theta_i)
\end{equation}

where $P_i = \prod_{j\neq i}p_j(\theta_j)$.

In this paper, we focus on symmetric Bayesian games, where all players share the same type space and type distribution.

\subsubsection{Transform Perfect Game to Imperfect Game with FHE}  
\label{sec:perfect2imperfect}
We propose a method to transform perfect information games into imperfect information games using Fully Homomorphic Encryption (FHE) techniques. The procedure is as follows:

\begin{enumerate}
  \item \textbf{Key Generation:} Each player $i \in \{1, \ldots, n\}$ generates a key pair using the BFV scheme's key generation algorithms:
    \begin{align}
      \texttt{sk}_i &\leftarrow \texttt{BFV.SecretKeyGen}(\lambda) 
    \end{align}
    where $\lambda$ is the security parameter.
  \item \textbf{Key Sharing:} Each player $i$ shares their secret key $\texttt{sk}_i$ using Shamir's secret sharing scheme (Section~\ref{sec:shamir}) to create a threshold key sharing scheme. This process can also be implemented using other DKG schemes like Pedersen's VSS\cite{fudenberg1991game} or Feldman's VSS\cite{feldman1987practical}\cite{gennaro2007secure}.
  
  \item \textbf{Strategy Encryption:} Each player encrypts their strategy $\sigma_i$ using the BFV scheme described in Section~\ref{sec:BFV}:
    \begin{equation}
      \sigma^\texttt{Enc}_i \leftarrow \texttt{BFV.Enc}(\texttt{pk}, \sigma_i)
    \end{equation}
  
  \item \textbf{Encrypted Strategy Distribution:} Each player distributes their encrypted strategy $\sigma^\texttt{Enc}_i$ to all other players.
  
  \item \textbf{Homomorphic Payoff Computation:} Each player $i$ computes their payoff function using homomorphic operations:
    \begin{equation}
      \pi^\texttt{Enc}_i \leftarrow f_i(\sigma^\texttt{Enc}_1, \ldots, \sigma^\texttt{Enc}_n)
    \end{equation}
    where $f_i$ is player $i$'s payoff function, computed in the encrypted domain using homomorphic operations.
  
  \item \textbf{Payoff Decryption and Proof Generation:} $t$ out of $n$ parties decrypts the ciphertext into plaintext by threshold decryption. Each player $i$ generates a zero-knowledge proof $\Pi_i$ to demonstrate the correctness of their payoff computation:
    \begin{equation}
      \Pi_i \leftarrow \texttt{Zk.Prove}(\texttt{sk}_i, \pi^\texttt{Enc}_i, \pi_i, \texttt{BFV.Dec})
    \end{equation}
    where $\pi_i$ is the decrypted payoff.
\end{enumerate}

\begin{algorithm}
  \caption{Transform Perfect Information Game to Imperfect Information Game with FHE}
  \label{alg:perfect2imperfect}
  \begin{algorithmic}[1]
  \Require Number of players $n$, Security parameter $\lambda$
  \Ensure Encrypted strategies $\{\sigma^\texttt{Enc}_i\}_{i=1}^n$, Payoffs $\{\pi_i\}_{i=1}^n$, Proofs $\{\Pi_i\}_{i=1}^n$
  
  \For{each player $i \in \{1, \ldots, n\}$ \textbf{in parallel}}
      \State $\texttt{sk}_i \gets \texttt{BFV.SecretKeyGen}(\lambda)$
      \State $\texttt{pk}_i \gets \texttt{BFV.PublicKeyGen}(\texttt{sk}_i)$
      \State $\sigma^\texttt{Enc}_i \gets \texttt{BFV.Enc}(\texttt{pk}_i, \sigma_i)$
      \State Broadcast $\sigma^\texttt{Enc}_i$ to all other players
  \EndFor
  
  \State Synchronization point: wait for all players to broadcast their encrypted strategies
  
  \For{each player $i \in \{1, \ldots, n\}$ \textbf{in parallel}}
      \State Receive $\{\sigma^\texttt{Enc}_j\}_{j \neq i}$ from other players
      \State $\pi^\texttt{Enc}_i \gets f_i(\sigma^\texttt{Enc}_1, \ldots, \sigma^\texttt{Enc}_n)$ \Comment{Homomorphic computation}
      \State $\pi_i \gets \texttt{BFV.Dec}(\texttt{sk}_i, \pi^\texttt{Enc}_i)$
      \State $\Pi_i \gets \texttt{Zk.Prove}(\texttt{sk}_i, \pi^\texttt{Enc}_i, \pi_i, \texttt{BFV.Dec})$
  \EndFor
  
  \State \Return $\{\sigma^\texttt{Enc}_i\}_{i=1}^n$, $\{\pi_i\}_{i=1}^n$, $\{\Pi_i\}_{i=1}^n$
  \end{algorithmic}
  \end{algorithm}

  \subsection{Voting Game Model}
  \label{sec:voting_game}
  In this chapter, we discuss how to use game theory in Monadring to design voting mechanisms to ensure the normal operation of the subnet. Here, we analyze the mathematical principles of different voting mechanisms. The significance lies in the fact that each AVS needs to design different voting mechanisms for different subnets. For voters, they hope to maximize their voting benefits. For the subnet, it hopes to distribute as few rewards as possible while ensuring that the entire network consensus is reached as soon as possible. We first introduced some basic game theory concepts, then applied these concepts to the subnets in Monadring. For the case of perfect information games, we provided a theoretically optimal voting strategy. For the case of imperfect information games, voters will try to find the right strategy, that is, $\xi$, to maximize their expected payoffs when reaching the Nash equilibrium. However, AVS designers hope to find the corresponding $\theta$ to minimize the overall payoff.
  \subsubsection{Voting Game}
  Assume there are $n$ voters, each with a voting weight of $\omega_i$, casting a vote $v\in{\{\top,\bot\}}$. The consensus threshold percentage is $\theta$, and their payoff function is:
  
  \begin{equation}
  b(\omega_i; \theta_\top,\theta_\bot, n) = 
  \begin{cases}
  \beta(\omega_i), & \text{if } \frac{\#(v = v_i)}{n} > \theta_{v_i} \\
  -\alpha(\omega_i), & \text{if } \frac{\#(v \neq v_i)}{n} > \theta_{\bar{v}_i} \\
  0, & \text{otherwise}
  \end{cases}
  \end{equation}
  where $n$ and $\theta$ refer to the number of participants and threshold percentage in this subgame, with $\theta_\top+\theta_\bot\geq 1$. We use $N$ and $\theta_g$ to denote the global threshold and number of participants. The $\beta(\omega_i)\geq 0$ and $\alpha(\omega_i)\geq 0$ are the payoff functions for a voter with a weight of $\omega_i$.
  
  We introduce a deterministic information: when voter $i$ votes, they know that there have already been $n^{(i)}\top$ votes for $\top$ and $n^{(i)}\bot$ votes for $\bot$. The threshold required for them in the subgame is:
  \begin{equation}
    \theta^{(i)}_\top = 
    \begin{cases}
      0, & \text{if } n^{(i)}_\top \geq \theta_\top\cdot n \\
      \infty, & \text{if } n^{(i)}_\bot \geq \theta_\bot\cdot n \\
      \frac{\theta_\top\cdot n - n^{(i)}_\top}{n - n^{(i)}}, & \text{otherwise}
    \end{cases}
  \end{equation}
  Therefore, at any point in time, each voter is essentially facing a subgame with parameters $\theta^i\top$, $\theta^i_\bot$, and $n^{(i)}$. In this game, the voter is completely unaware of the voting situation of the remaining $n-n^{(i)}-1$ individuals.
  \subsubsection{Perfect Information:$n^{(i)}=i-1$}
  \label{sec:perfect_information}
  Let's consider a special case where $n^{(i)}=i-1$, meaning that voter $i$ knows the voting behavior of all previous voters. Assuming all voters are rational, when the last voter votes, they know the voting behavior of all previous voters, and they only need to choose a vote that will satisfy the voting result.
  \begin{equation}
    \label{eq:rational_n}
    v^*_n = 
    \begin{cases}
      \top, &\text{if }n^n_\top+1>\theta_\top\cdot n\\
      \bot, &\text{if }n^n_\bot+1>\theta_\bot\cdot n\\
      \texttt{r}^{(n)}(\top,\bot) &\sim \mathcal{B}^{(n)}, \text{otherwise}
    \end{cases}
  \end{equation}
  Here, $v^*$ represents the rational voting result (not the actual vote), where $\mathcal{B}$ is the Bernoulli distribution.
  For the $n-1$ voters, if we assume they are rational, they would know the results of $n-2$ votes and $v^*_n$. Then, they can vote similarly to Eqn.~\ref{eq:rational_n} based on these $n-1$ results. Therefore, for any $i$, it can be inferred based on $\{v_1,v_2, \ldots,v_{i-1}\} \cup \{v^*_{i+1},v^*_{i+2},\ldots,v^*_n\}$.
  \begin{equation}
    \label{eq:rational_i}
    v^*_i = 
    \begin{cases}
      \top, &\text{if }n^{(i)}_\top+1+\sum_{j>i}{\mathbb{E}[v^*_j=\top]}>\theta_\top\cdot n\\
      \bot, &\text{if }n^{(i)}_\bot+1+\sum_{j>i}{\mathbb{E}[v^*_j=\bot]}>\theta_\bot\cdot n\\
      \texttt{r}^{(i)}(\top,\bot) &\sim \mathcal{B}^{(i)}(v;\theta_\top\cdot n-n^{(i)}_\top,\theta_\bot\cdot n-n^{(i)}_\bot,n-n^{(i)})
    \end{cases}
  \end{equation}
  If we assume that all other information for all voters is ideal, meaning there is no additional information that makes them more inclined to $\top$ or $\bot$, then $v^*$ is no longer a random variable, but:
  \begin{equation}
    \label{eq:rational_determined}
    \texttt{r}^{(i)} = 
    \begin{cases}
      \top, &\text{if }\theta_\top\cdot n-n^{(i)}_\top<\theta_\top\cdot n-n^{(i)}_\bot\\
      \bot, &\text{otherwise}
    \end{cases}
  \end{equation}
  However, the real situation is not like the ideal one, especially for the voters who are closer to the front ($i<\texttt{min}((1-\theta_\top)\cdot n+n_\top^{(i)},(1-\theta_\bot)\cdot n+n_\bot^{(i)})$). According to the Central Limit Theorem, the variance will be higher because they have to estimate the $\texttt{r}$ of other voters.
  \subsubsection{Imperfect Information: $n^{(i)}=0$}
  \label{sec:imperfect_information}
  For the case of $n^{(i)}=0$, where voter $i$ is the first voter and does not know the voting behavior of other voters, their voting result should be a random variable. This situation degenerates into a Bayesian game.
  Assume that its distribution is $v^*\sim B(X;\xi)$, where $B$ is the binomial distribution. For voter $i$, the expected payoff is:
  \begin{equation}
    \label{eq:expectation_top}
    \begin{aligned}
      &\mathbb{E}[b(\omega_i;\theta_\top,\theta_\bot,n )| v_i=\top]\\ 
      = &\beta\cdot\sum_{i>\theta_\top\cdot n-1}{\binom{n-1}{i}\xi^i(1-\xi)^{n-1-i}} - \alpha\cdot\sum_{i>\theta_\bot\cdot n}{\binom{n}{i}\xi^{(n-i)}(1-\xi)^{i}}
    \end{aligned}
  \end{equation}
  \begin{equation}
    \label{eq:expectation_bot}
    \begin{aligned}
      &\mathbb{E}[b(\omega_i;\theta_\top,\theta_\bot,n )| v_i=\bot]\\
       =& \beta\cdot\sum_{i>\theta_\bot\cdot n-1}{\binom{n-1}{i}\xi^{n-1-i}(1-\xi)^i}- \alpha\cdot\sum_{i>\theta_\top\cdot n}{\binom{n}{i}\xi^{i}(1-\xi)^{(n-i)}}
    \end{aligned}
  \end{equation}
  According to Bayes' theorem, the posteriori expectation is:
  \begin{equation}
    \label{eq:expectation_topl}
    \begin{aligned}
      &\mathbb{E}[b(\omega_i;\theta_\top,\theta_\bot,n)] \\
      = &\mathbb{E}[b(\omega_i;\theta_\top,\theta_\bot,n )| v_i=\bot]\cdot (1-\xi)+ \mathbb{E}[b(\omega_i;\theta_\top,\theta_\bot,n )| v_i=\top]\cdot \xi\\
    \end{aligned}
  \end{equation}
  We consider a Pareto optimal solution to Eqn.~\ref{eq:expectation_topl}, maximising the posterior expectation:
  \begin{equation}
    \label{eq:partial}
    \begin{aligned}
      &\frac{\partial\mathbb{E}[b(\omega_i;\theta_\top,\theta_\bot,n)]}{\partial\xi}\\
       =& \frac{\partial\mathbb{E}[b(\omega_i;\theta_\top,\theta_\bot,n )| v_i=\top]}{\partial\xi}\cdot \xi+ \mathbb{E}[b(\omega_i;\theta_\top,\theta_\bot,n )| v_i=\top]\\
      &+ \frac{\partial\mathbb{E}[b(\omega_i;\theta_\top,\theta_\bot,n )| v_i=\bot]}{\partial\xi}\cdot (1-\xi)-\mathbb{E}[b(\omega_i;\theta_\top,\theta_\bot,n )| v_i=\bot]\\
      & = 0
  \end{aligned}
  \end{equation}
  Deriving Eqn.~\ref{eq:expectation_top} and Eqn.~\ref{eq:expectation_bot} separately, we get:
  \begin{equation}
    \label{eq:expectation_top_partial}
    \begin{aligned}
      &\frac{\partial\mathbb{E}[b(\omega_i;\theta_\top,\theta_\bot,n )| v_i=\top]}{\partial\xi} \\
      =& \beta\cdot (n-1)\binom{n-2}{\theta_\top\cdot n-1}\int^{1-p}_0t^{(n - 1) - (\theta_\top\cdot n - 1) - 1} (1 - t)^{(\theta_\top\cdot n - 1)}\text{d} t\\
      &-\alpha\cdot n \cdot \binom{n-1}{\theta_\bot\cdot n}\int^{1-p}_0t^{n - (\theta_\bot\cdot n) - 1} (1 - t)^{\theta_\bot\cdot n}\\
      = &-\beta\cdot (n-1)\cdot \binom{n-2}{n\cdot\theta_\top-1}\xi^{n\cdot\theta_\top-1}(1-\xi)^{n\cdot(1-\theta_\top)-1} \\
      &+\alpha\cdot n \cdot \binom{n-1}{n\cdot\theta_\bot}\xi^{n\cdot\theta_\bot}(1-\xi)^{n\cdot(1-\theta_\bot)-1}\\
    \end{aligned}
  \end{equation}
  Similarly, we can find the partial derivative of $\mathbb{E}[b(\omega_i;\theta_\top,\theta_\bot,n )| v_i=\bot]$.
  \begin{equation}
    \label{eq:expectation_bot_partial}
    \begin{aligned}
      &\frac{\partial\mathbb{E}[b(\omega_i;\theta_\top,\theta_\bot,n )| v_i=\bot]}{\partial\xi} \\
      =& -\beta\cdot (n-1)\cdot \binom{n-2}{n\cdot\theta_\bot-1}\xi^{n\cdot\theta_\bot-1}(1-\xi)^{n\cdot(1-\theta_\bot)-1} \\
      &+\alpha\cdot n \cdot \binom{n-1}{n\cdot\theta_\top}\xi^{n\cdot\theta_\top}(1-\xi)^{n\cdot(1-\theta_\top)-1}\\
    \end{aligned}
  \end{equation}
  Solve Eqn.~\ref{eq:partial} by using Eqn.~\ref{eq:expectation_top_partial} and Eqn.~\ref{eq:expectation_bot_partial} to find the Nash equilibrium $\xi=\frac{\theta_\bot}{\theta_\top+\theta_\bot}$.
  With any parameters, $\xi=0$ and $\xi=1$ are the trivial Nash Equilibrium points. When $\alpha=\beta$,$\theta_\top=\theta_\bot$, the Nash equilibrium degenerates to $\xi=0.5$.
  
  \paragraph{Evaluation of $\theta$}
  For the designers of voting mechanisms, their goal is to achieve network-wide consensus at the minimum cost. That is, we should design $\theta, \beta, \alpha$ in such a way that consensus can be effectively reached at the lowest cost (note that these three are not necessarily constants, they could very well be in the form of functions such as $\beta(\theta)$ and $\alpha(\theta)$). Here we only discuss the design of $\theta$, while $\beta$ and $\alpha$ need to be evaluated based on functional analysis, which we will publish in the form of a paper. We derivate the expectation of the payoff function with respect to $\theta_\top$ and $\theta_\bot$. According to Eqn.~\ref{eq:expectation_top}, We approximately estimate that the likelihood expectation of the derivative of $\theta_\top\cdot n$ at the point $\theta_\top\cdot n -1$ is a Beta distribution. We use $\mathbb{E}_\top$ to abbreviate $\mathbb{E}[b(\omega_i;\theta_\top,\theta_\bot,n)|v_i=\top]$
  \begin{equation}
    \label{eq:expectation_top_to_top}
    \begin{aligned}
    \frac{\partial{\mathbb{E}}_\top}{\partial{\theta_\top}}=\frac{\partial\mathbb{E}_\top}{\partial(\theta_\top\cdot n)}\cdot\frac{\partial(\theta_\top\cdot n)}{\partial\theta_\top}= \beta\cdot n \cdot \text{f}_b(\xi;n\cdot\theta_\top,n-n\cdot\theta_\top+1)\cdot{n}\\
    \end{aligned}
  \end{equation}
  \begin{equation}
    \label{eq:expectation_top_to_bot}
    \begin{aligned}
    \frac{\partial{\mathbb{E}}_\top}{\partial{\theta_\bot}}= -\alpha\cdot n \cdot  \text{f}_b(1-\xi;\theta_\bot\cdot n+1,n-n\cdot\theta_\bot+1)\cdot(n+1)\\
    \end{aligned}
  \end{equation}
  \begin{equation}
    \label{eq:expectation_bot_to_bot}
    \begin{aligned}
    \frac{\partial{\mathbb{E}}_\bot}{\partial{\theta_\bot}}= \beta\cdot n \cdot \text{f}_b(\xi;n\cdot\theta_\bot,n-n\cdot\theta_\bot+1)\cdot{n}\\
    \end{aligned}
  \end{equation}
  \begin{equation}
    \label{eq:expectation_bot_to_top}
    \begin{aligned}
    \frac{\partial{\mathbb{E}}_\bot}{\partial{\theta_\top}}= -\alpha\cdot n \cdot  \text{f}_b(1-\xi;\theta_\top\cdot n+1,n-n\cdot\theta_\top+1)\cdot(n+1)\\
    \end{aligned}
  \end{equation}
  where $$f_b(x;a,b)=\frac{\Gamma(a+b)}{\Gamma(a)\Gamma(b)}x^{a-1}(1-x)^{b-1}$$ and $\Gamma(\cdot)$ is the gamma function.
  Then the condition for Bayesian Nash Equilibrium is:
  \begin{equation}
    \label{eq:partial_theta}
    \begin{aligned}
      \frac{\partial{\mathbb{E}}}{\partial{\theta_\top}}=\frac{\partial{\mathbb{E}_\top}}{\partial{\theta_\top}}\cdot\xi+\frac{\partial{\mathbb{E}_\bot}}{\partial{\theta_\top}}\cdot(1-\xi)= 0\\
      \frac{\partial{\mathbb{E}}}{\partial{\theta_\bot}}=\frac{\partial{\mathbb{E}_\top}}{\partial{\theta_\bot}}\cdot\xi+\frac{\partial{\mathbb{E}_\bot}}{\partial{\theta_\bot}}\cdot(1-\xi)= 0\\
    \end{aligned}
  \end{equation}
  Bringing in Eqn.~\ref{eq:expectation_top_to_top}, Eqn.~\ref{eq:expectation_top_to_bot}, Eqn.~\ref{eq:expectation_bot_to_bot}, Eqn.~\ref{eq:expectation_bot_to_top}, we get the analytic solution of $\theta_\top$ and $\theta_\bot$. But generally this analytical solution is not easy to solve, we can solve it by numerical methods.
  
  \paragraph{Regularization Term}
  To implement a reasonable voting mechanism, we need to introduce a regularization term. The purpose of this term is to prevent voters from voting too extremely, that is, their voting behavior is overly concentrated on $\top$ or $\bot$. By introducing a regularization term, we can encourage a more balanced voting behavior among voters. We can define the regularization term as follows:
  \begin{equation}
    \label{eq:regularization}
    \begin{aligned}
      \mathcal{F}(\xi, \theta,n,\alpha,\beta)=\mathbb{E}[b(\omega_i;\theta_\top,\theta_\bot,n)] + \underbrace{\lambda\cdot\left\|\frac{\partial{\mathbb{E}}}{\partial{\theta_\top}}+\frac{\partial{\mathbb{E}}}{\partial{\theta_\bot}}\right\|^2}_{\text{Regularization}}\\+\underbrace{\lambda_1\rho(B(X\leq\theta_\top;\xi))+\lambda_2\rho(B(X\leq\theta_\bot;\xi))}_{\text{Invalid Voting Mechanism Penalty}}\\
    \end{aligned}
  \end{equation}
  Where $\rho(\cdot)$ is a step function, similar to activation. When there is a high probability that the vote can be passed, the value is 0; otherwise, the value is 1. $\lambda$ is the regularization coefficient. We can find the optimal $\theta$, $\alpha$, and $\beta$ by minimizing $\mathcal{F}(\theta,n,\gamma,\beta)$.
  Use Eqn.\ref{eq:partial_theta} and Eqn.\ref{eq:partial} to find the optimal $\theta$ and $\xi$.

  Then the optimal $\theta$ and $\xi$ can be obtained by computing the minimum (for $\theta$) or the maximum (for $\xi$) of $\mathcal{F}$. Here we can use gradient descent method to solve it. We can solve $\theta$ and $\xi$ by iteration. In each iteration, we can update $\theta$ and $\xi$ by calculating $\frac{\partial{\mathcal{F}}}{\partial{\theta}}$ and $\frac{\partial{\mathcal{F}}}{\partial{\xi}}$. During the iteration process, we can determine whether there is convergence by calculating the value of $\mathcal{F}$. When the value of $\mathcal{F}$ converges, we can get the optimal $\theta$ and $\xi$.

  \subsubsection{Semi-Perfect Information: $0<n^{(i)}<i-1$}
  It is worth noting that in general, $\theta_\top$ and $\theta_\bot$ are the same. However, in the typical voting process, especially regarding blockchain consensus schemes, the situation often lies between a perfect information game and an imperfect information game. We refer to this as a semi-perfect game. For instance, in a consensus network, some nodes have already seen part of the voting results before voting, while they haven't seen the rest. After excluding the results they have seen, a voting subgame forms, which is an imperfect information game. And there is a high probability that $\theta_\top \neq \theta_\bot$. In this case, we can refer to the method described in Section~\ref{sec:imperfect_information} to characterize this subgame problem.

\section{Monadring Protocol}
\label{sec:protocol}
We want to formalize the procedure of organizing some partial nodes of an existing blockchain network to form a smaller subnet, i.e., implementing this protocol as a plugin of some specific blockchain node programs.
\subsection{Network Architecture}
\subsubsection{Model and definitions}
\paragraph{Hostnet and subnet.}
Consider a network composed of a set of participants $\mathbb{V}$ in which the majority obey a protocol to reach Byzantine Agreement and finality\cite{grandpa} over ledger $\mathcal{L}$.
$\mathcal{S}_{i}$ is a subset of $\mathcal{V}$, \( \mathcal{V} = \mathcal{S}_{0} \cup \mathcal{S}_{1} \cup .. \cup \mathcal{S}_{n-1}  \).
We call $\mathcal{V}$ is a \texttt{hostnet} and $\mathcal{S}_{i}$ is a \texttt{subnet} of $\mathcal{V}$.
\begin{itemize}
\item A node participant \( v \in \mathcal{V} \) could be a member of any $\mathcal{S}_{i}$ at meantime.\( v \in \mathcal{V}, v \in \mathcal{S}_{m} \cap \mathcal{S}_{n} \) is valid.
\item We assume the ledger $\mathcal{L}$ of hostnet maintains all the subnets information in form of a mapping \texttt{subnetid}\( \rightarrow \)\texttt{nodelist}.
\item Any participant of the hostnet could join a specific subnet through proposing a modification over ledger $\mathcal{L}$.
\item Since the ledger $\mathcal{L}$ is under Byzantine Agreement by all participants, the map can be considered a provable information outside any subnets.
\end{itemize}

\paragraph{Subnet ledger.}
For each subnet $\mathcal{S}_{i}$, all participants \(v_{j} \in \mathcal{S}_{i}\) maintain an independent ledger $\mathcal{L}_{i}$ different from the ledger $\mathcal{L}$ of hostnet, while the root state of each subnet ledger will be recorded in the hostnet ledger, \( \mathcal{L}_{i} \nsubseteq \mathcal{L}, \mathcal{F}(L_{i}) \in \mathcal{L} \).

A subnet could handle query and update requests independently from the hostnet ledger.
We can assume that the subnet ledger has properties as below:

\begin{itemize}
\item The subnet ledger contains all the modification events and each event has an incremental number as index.
The ledger is expected to be in a deterministic state after the $n_{th}$ event being applied, \(\mathcal{S}_{n+1} = f(\mathcal{S}_{n}, e_{n})\).
\item Particularly, changing the function $f$ is also a kind of event. The first event $e_{0}$ is loading the function.
\item The time complexity of looking up the $n_{th}$ event is $\mathcal{O}(1)$.
\item The time complexity of retrieving the maximum event id is $\mathcal{O}(1)$.
\end{itemize}

\paragraph{Token.} A \texttt{token} $\mathcal{T}$ of Monadring is a special signal circulates within a subnet.
A node in a subnet can write the subnet ledger only when it holds the token $\mathcal{T}$. The token is similar to a write lock in distributed networks.

\subsubsection{Subnet topology}
\label{sec:subnet_topology}
The participants of the subnet strictly follow the sequence of the \texttt{nodelist} to form a ring topology.
The \texttt{token} $\mathcal{T}$ of a subnet with $n$ nodes circulates around the subnet following the sequence of the \texttt{nodelist}.

\begin{itemize}
\item The \texttt{token} $\mathcal{T}$ of a subnet carries groups of events noted as $\mathbf{G}$ from its sender and the sender{'}s forehead recursively.
\item Group $\mathcal{G}_{i}$ is composed by the node $v_{i}$. It contains a list of modification events originally from its pending request queue, the node{'}s digital signature, a digest of its local ledger after these modifications applied, a number $q$ indicates how many times this group should delivered, the \texttt{nonce} of the signer and \texttt{ct} represents the voting data using FHE:
\begin{equation}
\mathcal{G}_{i} = (\mathbf{E} = [e_{k}, e_{k+1}..e_{k+n}], \mathcal{S}_{k+n}, \texttt{signature}, \texttt{nonce}, q, [\texttt{ct},..])
\end{equation}
Where the $\mathbf{E}$ could be empty, the $q$ could be negative, the \(\texttt{signature} = \texttt{sig}(\texttt{nonce}, \mathbf{E}, S_{k+n}, \texttt{nodekey})\).

\item Normally, a token circulates in a subnet with $n$ nodes should always include $n$ groups unless there were malicious behaviors or some nodes went offline.
Whenever a node receives the token, it ought to check the signature for each group.
Then applying all the events of each group and compare the digest with the local ledger.
The $q$ of executed groups should be decreased by 1.

\item If all checks pass, the node should handle transactions from its local queue and compose them as a new group with initial $q=n-1$ to replace previous one in the token.
Then try to deliver the token to its successor.
\end{itemize}

Assume a subnet with $n$ nodes \( v_{0}, v_{1} .. v_{n-1}, n\geq 3\). If no nodes joined the subnet during last round, when a node $v_{i}$ receives the token $\mathcal{T}$, noted as $\mathcal{T}_{i}$.
\begin{equation}
\mathcal{T}_{i} = \mathcal{G}_{i}, \mathcal{G}_{(i + 1)\mod n},.., \mathcal{G}_{(i + n - 1)\mod n}
\end{equation}
The $\mathcal{G}_{i}$ is the group generated by $v_{i}$ last time. After executed all the groups except the $\mathcal{G}_{i}$, $v_{i}$ should replace $\mathcal{G}_{i}$ with $\mathcal{G}_{i}^{\prime}$ and put it at the tail of the token, i.e., the $\mathcal{T}_{i+1}$ is:
\begin{equation}
\mathcal{T}_{i+1} = \mathcal{G}_{i+1}, \mathcal{G}_{(i+2)\mod n},.., \mathcal{G}_{i}^{\prime}; i+1<n
\end{equation}

\paragraph{Optimization.} A node could broadcast the events it just executed to others as an optimization.
The broadcasting doesn{'}t have to be through reliable communication since the token contains all the essential information that if some nodes don{'}t receive the broadcasting could execute it later.

A node could execute the events from broadcasting only if the \texttt{eventid} strictly follows the sequence in case it missed the previous broadcasting, e.g., a node has maximum event $e_{i}$ can pre-commit a group received from broadcasting in which the events start from $e_{i+1}$.

\subsection{Procedures}
\subsubsection{Launching a subnet and rotating subnet members}
\label{sec:launch_subnet_and_rotate}

To launch a subnet, a publishing procedure similar to deploying a smart contract is required. This involves a deterministic function $f$ in the form of executable binary code. Once the transaction is confirmed on the host network, any nodes can register to join the subnet. When the number of registered nodes reaches a certain \texttt{threshold}, these nodes will synchronize the function $f$ specified in the registration information and begin executing it locally. The token $\mathcal{T}$ initializes following the order of member registration, and the \texttt{FHEKeyGen} procedure mentioned previously will be executed.

After surpassing the threshold for registered nodes, subnet nodes must undergo periodic rotation to ensure randomness and reduce the potential for collusion. This rotation process can utilize a verifiable random function (VRF) or a simple pseudorandom function with a deterministic seed, such as the block hash at the end of each epoch, to select nodes for rotation in and out of the subnet.


\subsubsection{Node offline and token recovery}
The token might be disappeared during circulates within the subnet if some nodes go offline or due to malicious behaviors.

Assume a node $v_{i}$ of a subnet $\mathbf{S} = [v_{0}, v_{1}, .., v_{n-1}]$ holds the token $\mathcal{T}$ at moment $t$.
At the same time, all the rest nodes $\forall v_{j} \in \mathbf{S}, (i \ne j)$ keep the copy $\tau_{j}$ of the token $\mathcal{T}$ when they received it, e.g., $v_{0}$ keeps a copy $\tau_{0}$ of $\mathcal{T}_{0}$ delivered from $v_{n-1}$.
We call $\tau_{j}$ is the \textit{last seen token} of $v_{j}$.
The $\tau_{j}$ could be \texttt{null} since $v_{j}(j > i)$ may be a new member and never received the token before.

To recover the missing token, all nodes $v_{i} \in \mathbf{S}$ need to set a timer with countdown $\texttt{len}(\mathbf{S}) * \epsilon$ after they delivered it, where $\epsilon$ is the rough estimation of the execution time on a single node or a constant setting of the subnet.
When the timer of $v_{i}$ triggered, $v_{i}$ should resend the copy $\tau_{i}$ to its successor and reset the timer.
For a node $v_{i}$ receives a copy $\tau_{i}$ of the token, it is clearly to recognize that $\tau_{i}$ is a copy since $v_{i}$ had handled the events within it already.

\subsubsection{Function upgrade}
A function upgrade is an event that should be included in the token $\mathcal{G}$. Only specific users can initiate this event, where the subnet manager submits a transaction to modify the hostnet ledger $\mathcal{L}$, similar to registering a subnet. The transaction should include not only the function itself but also a digest and version, such as an executable WASM binary with its corresponding digest and version.

Once an upgrade transaction is confirmed by the hostnet, any members of the subnet can include this event into token $\mathcal{T}$ if the hostnet ledger contains a higher version of the function but the associated event is absent in the token.

\subsection{Fault Tolerance with FHE}
\label{sec:fault_tolerance}
The hostnet can reach the Byzantine Agreement over $\mathcal{L}$ based on an assumption that the ratio of honest participants $r > \texttt{threshold}$.
We don{'}t expect that a subnet has a same ratio as the honest participants of hostnet, neither we expect that a node was honest on ledger $\mathcal{L}$ would be honest on subnet legder $\mathcal{L}_{i}$.

We want to find a solution to detect the malicious behaviours or nodes offline then exclude them from the subnet by a provable invalidity.

For simplicity, we note $\mathcal{T}_{-i}$ as $\mathcal{T}_{i}$ excluding the $i_{th}$ group. An valid $\mathcal{T}_{i}$ must satisfy properties as below:

\begin{itemize}
\item The $q$ of $\mathcal{G}_{i}$ satisfies \(q \leq 0\). $q=0$ implies no new nodes joined the subnet in this round, while $q < 0$ implies some new nodes have joined the subnet. The node $v_{i}$ can check the \texttt{nodelist} from $\mathcal{L}$.
For rest $q$ of $\forall \mathcal{G} \in \mathcal{T}_{-i}$ are similar.
\item Group $\mathcal{G}_{i} \in \mathcal{T}$ contains a signature using its \texttt{nodekey}.
\item Assume the maximum event id of $\mathcal{G}_{i}$ is $m$. Combine all event lists of $\mathcal{T}_{-i}$ as a single list $\mathbf{E}_{-i}$. Assume the minimal event id of $\mathbf{E}_{-i}$ is $n$.
\(n = m+1\) and $m$ is the maximum event id of node $v_{i}$.
\end{itemize}

Given a token $\mathcal{T}$, any nodes can simply validate it.
If there is a malicious behavior noted as \( \mathcal{G}_{i} = (.., \mathbf{E} = [e_{i}, e_{i+1}, ..], ..) \), the nearest honest successor could place its event group with the smallest common agreed \texttt{eventid}, i.e., there are two conflicted groups $G_{i}^{\prime}$ and $\mathcal{G}_{i}$ in a token $\mathcal{T}$.
The invalid one contains the node's signature could be used for slashing.

The difference between this procedure and other BFT consensus algorithm is that the former one involves FHE to hide the voting information on each group.
Thus, the subnet could efficiently adapt to small-scale networks and avoid collusion actions.

\subsection{Framework Workflow}
\label{sec:framework_workflow}
The Framework Workflow of Monadring is described in Algorithm~\ref{alg:monadring}.
\begin{algorithm}[h!]
	\caption{Monadring Workflow}
  \label{alg:monadring}
	\begin{algorithmic}[1]
		\State Intialization: $sk, pk \leftarrow \texttt{FHEKeyGen}()$
    \State Sends a transaction to the hostnet $\mathcal{V}$ to opt-in a subnet $\mathcal{S}_i$.
    \State Ultilize \texttt{VRF} using Voting Model described in Section~\ref{sec:voting_game} to select the subnet members.
		\For{each epoch}
      \State Sharing/Resharing the keys using Shamir's Secret Sharing Scheme.
      \State Compose a new group with initial $q\leftarrow n-1$.
      \While {$q>0$}
        \If {receives $\mathcal{T}$}
          \State Check the signature of each group $\mathcal{G}_i$.
          \State Apply the events of each group $\mathcal{G}_i$.
          \State Compare the digest with $\mathcal{L}_i$.
          \State Decrease the $q$ of executed groups.
          \State Handle transactions from the local queue.
          \State Deliver $\mathcal{T}$ to its successor.
        \EndIf
          \State $q \leftarrow q - 1$
      \EndWhile
      \State Rotate the subnet members described in Section~\ref{sec:launch_subnet_and_rotate}.
      \If {Malicious behaviours detected}
        \State ApplyFault tolerance with FHE described in Section~\ref{sec:fault_tolerance}.
        \State Voting for slashing using the voting model described in Section~\ref{sec:voting_game}.
      \EndIf
		\EndFor
	\end{algorithmic}
  \end{algorithm}
\section{Future Work}
We are currently in the process of implementing several key solutions to advance our research:

\begin{itemize}
  \item Integration of Zama\cite{TFHE-rs}'s FHE library: We are working on incorporating Fully Homomorphic Encryption techniques using Zama's library to enhance privacy and security in our system.
  \item Adoption of Substrate framework: We are actively implementing the Substrate blockchain development framework to improve the scalability and interoperability of our solution.
  \item Decoupling of runtime and consensus: Our team is focusing on separating the runtime and consensus mechanisms, a change we anticipate will significantly enhance both flexibility and performance.
  \item Additional improvements: We are addressing numerous other areas for enhancement, including optimizing data structures, improving network protocols, and refining our cryptographic approaches.
\end{itemize}
These initiatives represent our current focus and the direction of our ongoing work. We will continue to refine and expand upon this paper over the coming months, providing more detailed analyses, implementation specifics, and preliminary results as our work progresses.

\bibliographystyle{unsrt}
\bibliography{monadring}

\begin{thebibliography}{10}

\bibitem{cryptoeprint:2016/889}
Aggelos Kiayias, Alexander Russell, Bernardo David, and Roman Oliynykov.
\newblock Ouroboros: A provably secure proof-of-stake blockchain protocol.
\newblock Cryptology ePrint Archive, Paper 2016/889, 2016.
\newblock \url{https://eprint.iacr.org/2016/889}.

\bibitem{burdges2020overview}
Jeff Burdges, Alfonso Cevallos, Peter Czaban, Rob Habermeier, Syed Hosseini, Fabio Lama, Handan~Kilinc Alper, Ximin Luo, Fatemeh Shirazi, Alistair Stewart, et~al.
\newblock Overview of polkadot and its design considerations.
\newblock {\em arXiv preprint arXiv:2005.13456}, 2020.

\bibitem{buchman2016tendermint}
Ethan Buchman.
\newblock {\em Tendermint: Byzantine fault tolerance in the age of blockchains}.
\newblock PhD thesis, University of Guelph, 2016.

\bibitem{acar2018survey}
Abbas Acar, Hidayet Aksu, A~Selcuk Uluagac, and Mauro Conti.
\newblock A survey on homomorphic encryption schemes: Theory and implementation.
\newblock {\em ACM Computing Surveys (Csur)}, 51(4):1--35, 2018.

\bibitem{chillotti2020tfhe}
Ilaria Chillotti, Nicolas Gama, Mariya Georgieva, and Malika Izabach{\`e}ne.
\newblock Tfhe: fast fully homomorphic encryption over the torus.
\newblock {\em Journal of Cryptology}, 33(1):34--91, 2020.

\bibitem{10.1007/978-3-642-32009-5_50}
Zvika Brakerski.
\newblock Fully homomorphic encryption without modulus switching from classical gapsvp.
\newblock In Reihaneh Safavi-Naini and Ran Canetti, editors, {\em Advances in Cryptology -- CRYPTO 2012}, pages 868--886, Berlin, Heidelberg, 2012. Springer Berlin Heidelberg.

\bibitem{cryptoeprint:2012/144}
Junfeng Fan and Frederik Vercauteren.
\newblock Somewhat practical fully homomorphic encryption.
\newblock Cryptology ePrint Archive, Paper 2012/144, 2012.
\newblock \url{https://eprint.iacr.org/2012/144}.

\bibitem{brakerski2014leveled}
Zvika Brakerski, Craig Gentry, and Vinod Vaikuntanathan.
\newblock (leveled) fully homomorphic encryption without bootstrapping.
\newblock {\em ACM Transactions on Computation Theory (TOCT)}, 6(3):1--36, 2014.

\bibitem{pang2005new}
Liao-Jun Pang and Yu-Min Wang.
\newblock A new (t, n) multi-secret sharing scheme based on shamir’s secret sharing.
\newblock {\em Applied Mathematics and Computation}, 167(2):840--848, 2005.

\bibitem{groth2021non}
Jens Groth.
\newblock Non-interactive distributed key generation and key resharing.
\newblock {\em Cryptology ePrint Archive}, 2021.

\bibitem{fudenberg1991game}
Drew Fudenberg and Jean Tirole.
\newblock {\em Game theory}.
\newblock MIT press, 1991.

\bibitem{daskalakis2009complexity}
Constantinos Daskalakis, Paul~W Goldberg, and Christos~H Papadimitriou.
\newblock The complexity of computing a nash equilibrium.
\newblock {\em Communications of the ACM}, 52(2):89--97, 2009.

\bibitem{feldman1987practical}
Paul Feldman.
\newblock A practical scheme for non-interactive verifiable secret sharing.
\newblock In {\em 28th Annual Symposium on Foundations of Computer Science (sfcs 1987)}, pages 427--438. IEEE, 1987.

\bibitem{gennaro2007secure}
Rosario Gennaro, Stanislaw Jarecki, Hugo Krawczyk, and Tal Rabin.
\newblock Secure distributed key generation for discrete-log based cryptosystems.
\newblock {\em Journal of Cryptology}, 20:51--83, 2007.

\bibitem{grandpa}
Eleftherios Kokoris-Kogia Alistair~Stewart.
\newblock Grandpa: a byzantine finality gadget.
\newblock {\em arXiv:2007.01560}, 2020.

\bibitem{TFHE-rs}
Zama.
\newblock {TFHE-rs: A Pure Rust Implementation of the TFHE Scheme for Boolean and Integer Arithmetics Over Encrypted Data}, 2022.
\newblock \url{https://github.com/zama-ai/tfhe-rs}.

\end{thebibliography}
\end{document}